\documentclass[twoside,journal]{IEEEtran}
\usepackage{blindtext}
\usepackage{graphicx}

\usepackage[T1]{fontenc}

\usepackage{cite}
\usepackage[cmex10]{amsmath}
\usepackage{hyperref}
\usepackage[mathscr]{euscript}

\usepackage{algorithm}
\usepackage{algorithmic}

\usepackage{color}
\usepackage[table]{xcolor}

\usepackage{amssymb}



\usepackage[font=small]{caption}
\usepackage{subcaption}

\usepackage{textcomp}

\hypersetup{
	colorlinks,
	citecolor=black,
	filecolor=black,
	linkcolor=black,
	urlcolor=black
}
\newtheorem{theorem}{Theorem}
\newtheorem{lemma}{Lemma}
\newtheorem{remark}{Remark}
\newtheorem{definition}{Definition}
\newtheorem{assumption}{Assumption}
\newtheorem{proposition}{Proposition}
\newtheorem{example}{Example}

\let\oldtheorem\theorem
\let\endoldtheorem\endtheorem
\def\theorem{\begingroup \oldtheorem }
\def\endtheorem{ \hfill $\square$\endoldtheorem \endgroup}

\let\oldlemma\lemma
\let\endoldlemma\endlemma
\def\lemma{\begingroup \oldlemma }
\def\endlemma{ \hfill $\square$\endoldlemma \endgroup}

\let\oldremark\remark
\let\endoldremark\endremark
\def\remark{\begingroup \oldremark }
\def\endremark{ \hfill $\square$\endoldremark\endgroup}

\let\oldproposition\proposition
\let\endoldproposition\endproposition
\def\proposition{\begingroup \oldproposition}
\def\endproposition{ \hfill $\square$\endoldproposition\endgroup}

\let\olddefinition\definition
\let\oldenddefinition\enddefinition
\def\definition{\begingroup \olddefinition  }
\def\enddefinition{ \hfill $\square$\oldenddefinition\endgroup}

\let\oldassumption\assumption
\let\oldendassumption\endassumption
\def\assumption{\begingroup \oldassumption }
\def\endassumption{ \hfill $\square$\oldendassumption\endgroup}

\let\oldexample\example
\let\endoldexample\endexample
\def\example{\begingroup \oldexample }
\def\endexample{ \hfill $\square$\endoldexample \endgroup}

\newcolumntype{P}[1]{>{\centering\arraybackslash}m{#1}}
\renewcommand\IEEEkeywordsname{Index Terms}

\let\oldIEEEkeywords\IEEEkeywords
\let\oldendIEEEkeywords\endIEEEkeywords
\def\IEEEkeywords{\begingroup \oldIEEEkeywords \begingroup\normalfont\bfseries}
\def\endIEEEkeywords{ \endgroup\oldendIEEEkeywords \endgroup}

\usepackage{lettrine}

\usepackage{cite}

\usepackage{soul}

\usepackage{enumerate}

\begin{document}
\IEEEpeerreviewmaketitle
\title{{ Leader-Follower Network Aggregative Game with Stochastic Agents' Communication and Activeness}}

\author{Mohammad~Shokri,
	and~Hamed~Kebriaei,~\IEEEmembership{Senior Member,~IEEE}
	\thanks{M. Shokri and H.Kebriaei are with the School of Electrical
		and Computer Engineering, College of Engineering, University of Tehran, Iran. (email:
		mo.shokri@ut.ac.ir, kebriaei@ut.ac.ir)}
}


\maketitle
\begin{abstract}
This technical note presents a leader-follower scheme for network aggregative games. The followers and leader are selfish cost minimizing agents. The cost function of each follower is affected by strategy of leader and aggregated strategies of its neighbors through a communication graph. The leader infinitely often wakes up and receives the aggregated strategy of the followers, updates its decision value and broadcasts it to all the followers. Then, the followers apply the updated strategy of the leader into their cost functions. { The establishment of information exchange between each neighboring pair of followers, and the activeness of each follower to update its decision at each iteration are both considered to be drawn from two arbitrary distributions.} Moreover, a distributed algorithm based on sub-gradient method is proposed for updating the strategies of leader and followers. The convergence of the proposed algorithm to the unique generalized Nash equilibrium point of the game is proven in { both almost sure and mean square} senses.
\end{abstract}
\section{Introduction}
Distributed optimization over networks has attracted widespread attention of researchers in recent years \cite{MAI2019}.  As a typical framework, each agent in a network aims to minimize a social  or an individual cost function while it communicates with some other agents through the network. In case that each agent is modeled as a selfish player who aims to minimize its own cost and also, the agent's cost is affected by decision variables of its neighbors through the network topology, the problem can be studied as a non-cooperative network game \cite{SALEHISADAGHIANI2018}. If the effect of decision variables of rivals on the agent's cost function appears as an aggregative term (e.g. summation or weighted sum), the network game is known as network aggregative game (NAG) \cite{Galeotti2010}. Many applications can be studied via this framework including, power system \cite{Zhao2014}, opinion dynamics \cite{opinion_dynamics_grammatico2017},  communication system \cite{Semasinghe2018}, provision of public goods \cite{Allouch2015} and criminal networks \cite{Chen2018}.   
\par
In a class of NAGs, the cost function of each agent is affected by the aggregated strategies of all network agents \cite{LIANG2017,Koshal2016,Paccagnan2018} (including neighbors and non-neighbors). In this case, the coupling term among the agents is the same for all of them. This assumption does not cover the problems that agents have different relevancy or limited communication  capabilities. Such challenges motivate studying another type of NAGs in which agents have their dependency with their neighbors. In \cite{Parise2015}, NAGs are studied for the agents with quadratic cost functions. Therein, some algorithms are proposed which converge to the Nash equilibrium point using the best response functions while at each iteration, agents communicate with their neighbors and update their strategies. Furthermore, in \cite{proximal_Grammatico_2018}, a Nash seeking dynamics is utilized for agents with proximal quadratic cost functions whose best responses are in the form of proximal operator.  In \cite{PARISE2019}, the cost function of the agents is considered in general form. In this work, a distributed algorithm has been utilized in which each agent communicates with all its neighbors at each iteration and updates its strategy based on its best response function. However, the algorithms are designed based on the best response scheme which imposes high computational complexity, specially for a general cost function. In such a case, as mentioned in \cite{Simon1996}, the players naturally dismiss strategies which are characterized by high computational cost and hence, the gradient response seems to an appropriate choice for updating the decision of the agents \cite{Yin2010}.  
\par
In the mentioned researches on NAGs, all the agents are in the same order of decision making with the same "structure" of cost functions and communication type. However, in many applications, there is a high-level agent (leader) who aims to optimize its own objective function which also depends on the strategies of other agents at a lower level (followers). Several researches have investigated leader-follower games \cite{Nourian2012,Kebriaei2018,Wang2014} and it has been extensively utilized in many engineering fields such as wireless sensor networks \cite{Saffar2017}, supply chain  management \cite{Yu2009}, and smart grid \cite{Elrahi2019}. If the leader has complete information about the followers' cost functions, then the concept of Stackelberg equilibrium can be applied directly \cite{Kebriaei2018}. In this case, the problem can be solved as a bi-level optimization in which the leader first computes the reaction function of the followers with respect to its strategy. Then, by applying the reaction function of the followers into its own cost function, the leader finds its optimal strategy. However, if the leader does not have a-priori information about the cost function of the followers, then the leader needs to learn its optimal strategy by iterative methods. In \cite{Nourian2012}, a leader-following problem is discussed in which the cost function of the leader is independent of the followers' strategies and only the followers respond to leaders' strategy. In \cite{Wang2014}, an iterative hierarchical mean-field game is studied including a leader and a large number of followers. The leader first announces its decision and then, followers respond by knowing the leader's decision.  
\par
In this paper, we propose a leader-follower NAG. The leader has a different type of cost function from the followers which is affected by the aggregated strategy of all the followers. Additionally, the leader has a different type of communication and activeness from the followers. It is considered that the leader infinitely often: wakes up, receives the last aggregated strategy of followers, updates its strategy, broadcasts it to all the followers, and then goes to sleep. In the lower level, the followers receive the last decision value of the leader and play a NAG until the next decision update of the leader. The cost function of each follower is affected by aggregated strategies of its neighbors and also the strategy of the leader. We also consider stochastic communication and activeness of the agents in NAG. At each arbitrary iteration, based on a stochastic binary distribution, each follower may become active to update its decision based on the projected sub-gradient method. Besides, at each iteration, an agent may receive the decision value of a neighboring agent based on another stochastic binary distribution. The corresponding stochastic binary variables of the two mentioned distributions can be dependent on each other, and further, there might be some constraints on those variables. Finally, a distributed algorithm is proposed in which the decision values of the leader and followers converge to the unique Generalized Nash Equilibrium (GNE) point of the game. 
\par
{  To the best of our knowledge, compared to single-level NAGs \cite{Parise2015,proximal_Grammatico_2018,PARISE2019}, this is the first paper that proposes a leader-follower scheme for NAGs. Further, this is the first paper that presents a general stochastic framework that simultaneously considers communication and activeness of the agents in NAGs in which the Gossip based communication protocol \cite{Boyd2006} can be encountered as a special case of the proposed framework. From other aspects, compared to aggregative games which consider the average strategy of whole population as a common coupling term among the agents \cite{LIANG2017,Koshal2016,Paccagnan2018}, in this paper, the local aggregative term is studied in which only neighbors of each follower, as well as the leader, affect the follower's cost function. Compared to the literature of NAGs, those consider the local aggregative term, we have studied a general strongly convex cost function instead of quadratic one \cite{Parise2015,proximal_Grammatico_2018}. In contrast with the papers which have utilized the best response dynamics as the agents' decision update rule \cite{Parise2015,proximal_Grammatico_2018,PARISE2019}, we have used the projected sub-gradient method for optimization to cope with the limited computational capabilities of the agents. The main contributions of this paper can be summarized as follow:   
\begin{itemize}
	\item We propose a leader-follower framework for NAG.
	\item We study stochastic communication and activeness of the agents in NAG.
	\item  A distributed algorithm based on projected sub-gradient method is proposed and its convergence to the unique GNE point of the game is proven in both { almost sure and mean square} senses.
\end{itemize}
}
\par 
This paper is structured as follows. The system model is introduced in Section \ref{system_model_section}. In Section \ref{LeaderFollower_network_game}, communications framework and information structure are given and a distributed optimization algorithm is proposed for decision making of the agents. The convergence of the algorithm to the GNE point of the game is proven in Section \ref{Convergence_Analysis_Section}. Simulation results are presented in Section \ref{Simulation_Results}. Finally, Section \ref{Conclusion_FutureWork_Section} summarizes the results and draws conclusion. 

\section*{Notation and Preliminaries}
$\mathbb{N}$ and $\mathbb{R}$ are the set of natural and real numbers, respectively. $|{\cal N}|$ denotes the number of members of the set ${\cal N}$. Let  $A^{\top}$ denotes the transpose of a vector/matrix $A$. $||A||$ indicates the matrix norm which is equal to the largest eigenvalue of the matrix. The 2-norm of vector $x$ is defined by $\left\|x\right\|=\sqrt{x^\top x}$.  $\textbf{col}(x_1,...,x_N)=[x_1^\top,...,x_N^\top]^\top$ indicates the column augmentation of column vectors $x_n$ for $n=1,...,N$. ${\vec 1}_n=\textbf{col}(1,...,1)$ and ${\vec 0}_n=\textbf{col}(0,...,0)$ where ${\vec 1}_n,{\vec 0}_n\in \mathbb{R}^n$. The probability function and expected value are denoted by $\textbf{P}\{.\}$ and $\textbf{E}\{.\}$, respectively. Supposing the function $f(.):{\cal X}\to \mathbb{R}$, $g(x^\prime)$ is called the sub-gradient of $f(.)$ at $x^\prime$ if $\forall x \in {\cal X}:f(x^\prime)+ (x-x^\prime)^\top g(x^\prime)\le f(x)$. Also, the projection operator of ${\cal X}$ is defined by $\Pi_{\cal X}(x)=\arg\min_{y\in{\cal X}}||y-x||^2$. { $g(x)$ is strictly monotone if $(g(x_2,r)-g(x_1,r))^\top (x_2-x_1)>0$ for $\forall x_1,x_2: x_1\ne x_2$.}

\section{System Model}
\label{system_model_section}
Consider a set of follower agents ${\cal N}=\{1,...,N\}$ and a leader involved in a non-cooperative game. The followers are connected to each other via a communication network represented by a directed graph $G(\cal N, \cal A)$ where ${\cal A}=[a_{nm}]_{n,m\in{\cal N}}$ is adjacency matrix of $G$  such that $a_{nm}=1$ if there is a communication link from follower $m$ to $n$, and $a_{nm}=0$ otherwise (more details on communication network of the followers is given in Section \ref{information_fellows}).  Each follower $n \in {\cal N}$ has its decision variable (i.e. strategy) $x_n \in {\cal X}_n$ where ${\cal X}_n \subset \mathbb{R}^{M^F}$ is a non-empty, compact and convex set. The cost function of  follower $n$ depends on the aggregated strategy of its neighbors whose set is denoted by ${\cal N}_n$, and also the strategy of the leader. Therefore, the cost function of follower $n$ is defined as follows
{\begin{equation}
	{J_n^F(x_n,\sigma_n(x_{-n}),y)}:x_n \in {\cal X}_n
	\label{followers_cost_function}
\end{equation}}
where  $x_{-n}=\textbf{col}(x_1,...,x_{n-1},x_{n+1},...,x_N)$, and $y \in {\cal Y}$ is the strategy of leader which is selected from a compact and convex set denoted by ${\cal Y}\subset \mathbb{R}^{M^L}$. { Also, we define $d_n(x_n,\sigma_n(x_{-n}),y)$ as a sub-gradient of $J_n^F(x_n,\sigma_n(x_{-n}),y))$ with respect to $x_n$.} $\sigma_n(x_{-n})$ is the aggregated strategy of follower-$n$'s neighbors which is defined as
\begin{equation}
	\begin{split}
		\sigma_n(x_{-n})&=\sum\limits_{m \in {\cal N}_n} {w_{nm} x_m}.\\
	\end{split}
	\label{follower_neighbor_aggregator}
\end{equation}
where $\sum\limits_{m \in {\cal N}-\{n\}}w_{nm}=1$, $w_{nm}> 0$ if $a_{nm}=1$, and $w_{nm}= 0$ if $a_{nm}=0$.  Hence, we can define the weight matrix of the graph $G$ by ${\cal W}=[w_{nm}]_{n,m\in{\cal N}}$. The cost function of the leader is defined as follows
{\begin{equation}
	{J^L(y,\sigma_{0}(x_{\cal N}))}:y \in {\cal Y}.
	\label{leader_cost_function}
\end{equation}}
{ Further, $d_0(y,x_{{\cal N}})$ denotes a sub-gradient of $J^L(y,\sigma_{0}(x_{\cal N}))$ with respect to $y$.}  The cost function of  leader is also affected by the aggregated strategy of the followers $\sigma_{0}(x_{\cal N})$ defined by
\begin{equation}
	\sigma_0(x_{\cal N})=\sum\limits_{n \in {\cal N}}w_{0n} { x_n}
	\label{leader_aggregator}
\end{equation}
where $x_{\cal N}=\textbf{col}(x_1,...,x_N)$, $w_{0n}$ denotes the weight of bidirectional communication link between follower $n$ and the leader, and we have $\sum_{n\in {\cal N}}w_{0n}=1$, and $w_{0n}\ge 0$. let ${\vec w}_0=\textbf{col}(w_{01},...,w_{0N})$ denotes the leader weight vector. 
\par
Accordingly, the non-cooperative game among the followers and leader is defined as follows
\begin{equation}
	\begin{split}
	{\cal G}=\left\{\begin{array}{l}
	\textbf{Players:}\text{ followers }{\cal N}\text{ and the leader}\\
	\textbf{Strategies:}\left\{\begin{array}{ll}
			\text{Follower }n\text{: }&x_n\in{\cal X}_n\\
			\text{Leader: }&y\in{\cal Y}\\
		\end{array}\right.\\
		\textbf{Cost:}\left\{\begin{array}{ll}
			\text{Follower }n\text{: }&J_n^F(x_n,\sigma_n(x_{-n}),y)\\
			\text{Leader: }&J^L(y,\sigma_{0}(x_{\cal N}))\\
		\end{array}\right.\\
	\end{array}\right.
	\end{split}
	\label{Game_equation}
\end{equation} 
\begin{assumption}
	$J_n^F(x_n,\sigma_n(x_{-n}),y)$ and $J^L(y,\sigma_{0}(x_{\cal N}))$ are sub-differentiable and strongly convex over ${\cal X}_n$ and ${\cal Y}$   with respect to $x_n$ and $y$, respectively. i.e. there exist $C_n$ and $C_0$ for $\forall n\in {\cal N}$ such that
	\begin{equation}
		\begin{split}
			&(d_n(x_n,\sigma,y)-d_n(x_n^\prime,\sigma,y))^\top(x_n-x_n^\prime)\ge C_n  ||x_n-x_n^\prime||^2\\
			&(d_0(y,\sigma)-d_0(y^\prime,\sigma))^\top(y-y^\prime)\ge C_0  ||y-y^\prime||^2\\
		\end{split}
		\label{strong_convex_inequality}
	\end{equation}
	Also, there exist Lipschitz constants $L$ and  $L_{0}$ such that  $\forall n \in {\cal N},\forall x_n\in{\cal X}_n,y\in {\cal Y},\forall \sigma_1,\sigma_2 \in \{\sigma_n(x_{-n})|x_{-n} \in \prod_{m\ne n} {\cal X}_m\}$, we have
	\begin{equation}
		\begin{split}
			||d_n(x_n,\sigma_1,y_1)-d_n(x_n,\sigma_2,y_2)||&\le L  ||\sigma_1-\sigma_2||\\
			&+L  ||y_1-y_2||\\
			||d_0(y,\sigma_1)-d_0(y,\sigma_2)||&\le L_0  ||\sigma_1-\sigma_2||
		\end{split}
		\label{lipschitz_assumption_equation}
	\end{equation} 
	where $d_n(x_n,\sigma_n(x_{-n}),y)$ and  $d_0(y,\sigma_n(x_{-n}))$ are the sub-gradients of  $J_n^F$ and $J^L$, respectively.
	\label{set_and_function_assumption}
\end{assumption}
\par
The equilibrium point of the aforementioned leader-follower game is defined as follows:
\begin{definition}
	$(x_{\cal N}^*,y^*)$ is  a Generalized Nash equilibrium (GNE) point   of the leader-follower game between the followers and the leader if
	{ \begin{equation}
		\begin{split}
			 \forall x_n \in {\cal X}_n:&  J_n^F(x_n^* ,\sigma(x_{-n}^*),y^*)\le J_n^F(x_n ,\sigma(x_{-n}^*),y^*)\\
			 \forall y \in {\cal Y}:& J^L(y^*,\sigma_0(x_{\cal N}^*))  \le J^L(y,\sigma_0(x_{\cal N}^*))
		\end{split}
		\label{GNE_equation}
	\end{equation}}
	$\forall n \in {\cal N}$ where  $x_{-n}^*=\textbf{col}(x_1^*,...,x_{n-1}^*,x_{n+1}^*,...,x_N^*)$ and $x_{\cal N}^*=\textbf{col}(x_1^*,...,x_N^*)$.
	\label{GNE_definition}
\end{definition}

\section{The Leader-Follower Network Game}
\label{LeaderFollower_network_game}
\subsection{Communication and Information Structure} \label{information_fellows}
{ 1) Followers' Communication and Activeness: We consider that follower $n$ receives information of its neighbor $m \in {\cal N}_n$ at iteration $k$ with probability $p_{mn}^k$. Furthermore, the follower $n$ is active at iteration $k$ to update its decision with probability $q_{n}^k$. Let random binary variables $l_{n,m}^k$ and $e_n^k$ denote the establishment of communication from follower $n$ to $m$, and activeness of follower $n$ at iteration $k$, respectively. Clearly, $\forall k \ge 0 :l_{n,m}^k=0$ for non-neighbor followers ($a_{nm}=0$).  Then, the last  information of follower $n$ from follower $m$ denoted by $\tilde{x}_{n,m}^k$ is updated as follows:}
\begin{equation}
\begin{split}
\tilde{x}_{n,m}^{k+1}&=(1-l_{n,m}^k) \tilde{x}_{n,m}^{k} +l_{n,m}^k x_m^k.\\
\end{split}
\label{latest_data_follower}
\end{equation}
 Based on (\ref{latest_data_follower}), the  aggregated strategy of neighborhoods of follower $n$ at iteration $k$ is calculated as $\tilde{\sigma}_n^k=\sigma_n(\tilde{x}_{n,-n}^k)$ where $\tilde{x}_{n,-n}^k=\textbf{col}(\tilde{x}_{n,1}^k,...,\tilde{x}_{n,n-1}^k,\tilde{x}_{n,n+1}^k,...,\tilde{x}_{n,N}^k)$.
\par
Let's consider ${\cal L}^k=[l_{nm}]_{n,m\in {\cal N}}$ and ${\cal E}^k=\textbf{col}(e_1^k,...,e_N^k)$ as the connectivity matrix and the activity vector of the followers at iteration $k$, respectively. { The constraint set ${\cal P}$ represents the set from which ${\cal L}^k$ and ${\cal E}^k$ are selected for all iterations $\forall k \ge 0$.} Further, ${\cal H}^k$ denotes the history set of stochastic variables ${\cal L}^k$ and ${\cal E}^k$ up to iteration $k$  which is defined by ${\cal H}^{k+1}={\cal H}^k \cup \{({\cal L}^k,{\cal E}^k)\}$ and  ${\cal H}^{1}=\{({\cal L}^0,{\cal E}^0)\}$. In this paper, the probabilities of ${\cal L}^k$ and ${\cal E}^k$ are considered to be possibly dependent to ${\cal H}^{k}$ and  ${\cal P}$ as follows:
\begin{equation}
\begin{split}
\textbf{P}\{l_{nm}^k=1|{\cal F}^{k}\}=p_{nm}^k,\textbf{P}\{e_{n}^k=1|{\cal F}^{k}\}=q_{n}^k
\end{split}
\label{probability_of_variables}
\end{equation}
where ${\cal F}^{k} = {\cal H}^{k} \cap {\cal P}$. Apparently, ${\cal F}_k \subset {\cal F}_{k+1}$ holds for $\forall k \ge 0$. To illustrate the  dependency among stochastic variables, in what follows, the well-known Gossip-based communication protocol \cite{Boyd2006} has been studied as an example of the  proposed communication framework. 
\begin{example}[Gossip-based Communication]
	Suppose that the followers communicate to each other through an undirected  graph $G(\cal N, \cal A)$. In gossip-based communication, each node has an independent stochastic clock which ticks with rate 1 Poisson process  and makes the corresponding agent become active. The simultaneous clock ticks are neglected and as a result, at most one agent wakes up at each time slot of global clock. For instance, at $k^{th}$ time slot, let the clock of agent $n$ ticks. It wakes up and contacts with only one neighbor (say agent $m$). They communicate with each other, update their strategies and then both go to sleep.  In this case, our communication framework imposes the following constraints:
	\begin{equation}
	\begin{split}
	{\cal P}=\bigg\{&({\cal L}^k,{\cal E}^k)\bigg|\sum_{n\in {\cal N}}e_n^k= 2,l_{nm}^k= e_n^k e_m^k,\\
	&e_n^k e_m^k\le a_{nm},\forall n,m \in {\cal N} \bigg\}
	\end{split}
	\label{gossip_constraint}
	\end{equation}
	The first constraint indicates that only two player could be active in iteration $k$. $l_{nm}^k= e_n^k e_m^k$ implies that a link is established when both of its sender and receiver are active.  In addition, $l_{nm}^k= l_{mn}^k$ can be concluded from this constraint. Furthermore, inequality constraint $e_n^k e_m^k\le a_{nm}$ prevents two non-neighbors to become active. The probability of each link and each node are calculated as follows:
	\begin{equation}
	\begin{split}
		p_{nm}^k=\frac{1}{N}(\frac{1}{|{\cal N}_n|} + \frac{1}{|{\cal N}_m|}),q_{n}^k=\frac{1}{N}(1 + \sum_{m\in {\cal N}_n}\frac{1}{|{\cal N}_m|})
	\end{split}
	\label{gossip_probability}
	\end{equation}
	where ${\cal N}_n=\{m|a_{mn}=1\}$.
	\label{gossip_example}
\end{example}
\begin{assumption}
	There exist $\gamma>0$ and $\delta>0$ such that ${p}_{nm}\ge\gamma$ and ${q}_{n}\ge\delta$ for $\forall n\in {\cal N}, \forall m \in {\cal N}_m$.
	\label{probabilities_assumption}
\end{assumption}
Note that ${p}_{nm}=0$ for $\forall n\in {\cal N},\forall m \notin {\cal N}_m$. 

\par
2) Communication between leader and followers: It is assumed that the leader does not exchange information with the followers at every iterations. Instead, it is considered that the leader infinitely often wakes up and receives aggregated strategy of the followers in an  arbitrary desired iteration set ${\cal K}^L=\{k^L_i\}_{i=0}^{\infty}$. The leader also updates its decision variable and broadcasts it to all the followers at the same iteration. It is supposed that $k^L_0=0$.
\begin{assumption}
	There is ${\bar K}<\infty$ such that $k^L_{i+1}-k^L_i\le{\bar K}$ for $\forall i \in \mathbb{N}$.
	\label{leader_max_iteration_lenght}
\end{assumption} 

\begin{figure}[t!]
 	\centering
 	\includegraphics[width=7cm,height=5cm]{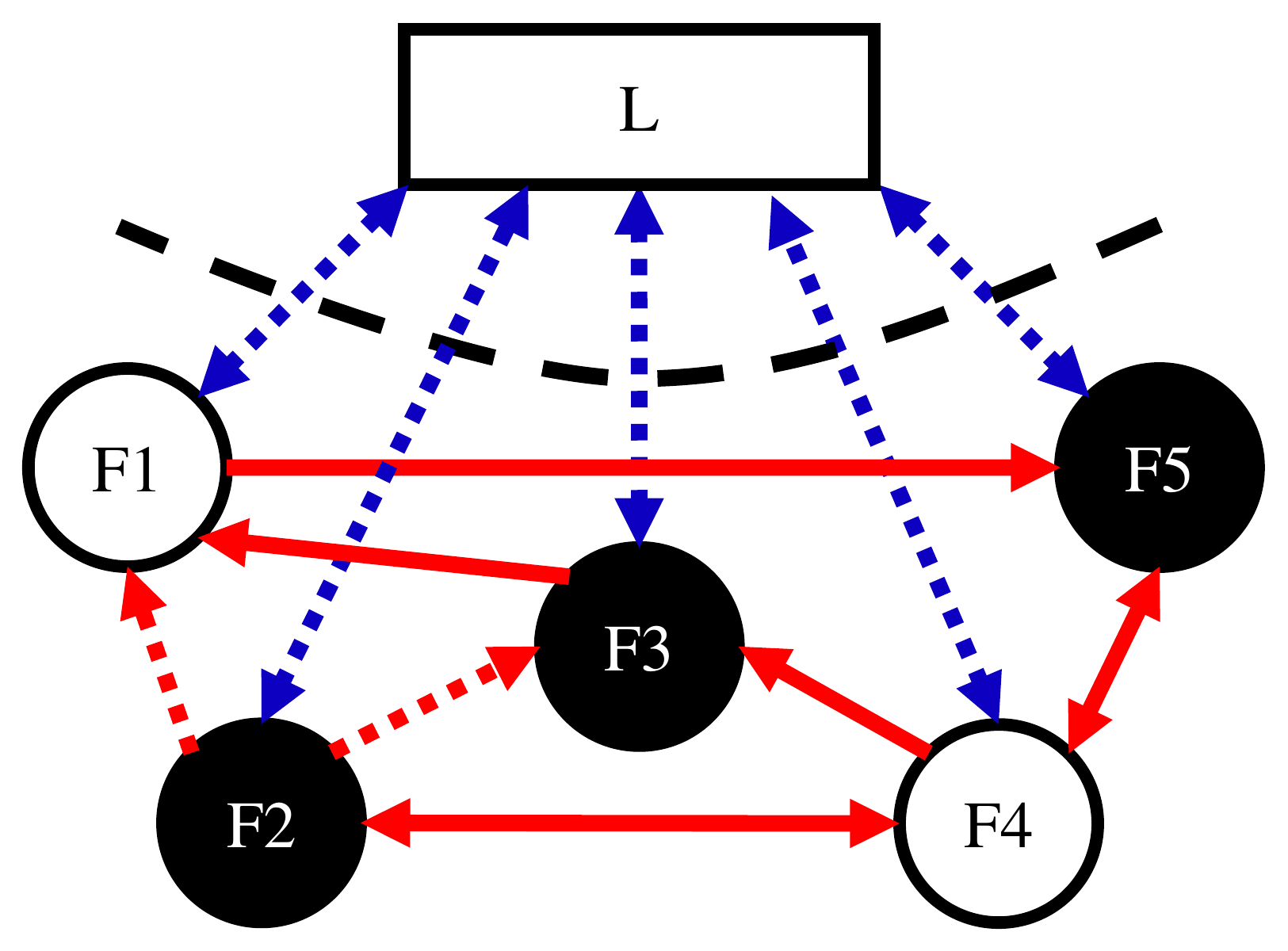}
 	\caption{The information scheme of  leader-follower network game between followers ("F" shapes) and the leader ("L" shape). The filled and unfilled objects indicate the active and inactive agents to make decision, respectively. The dashed line indicates that there exists a communication line, but it's not establish at iteration $k$.}
 	\label{information_scheme}
 \end{figure}

 The schematic of information flows among followers and between followers and leader is shown in Fig. \ref{information_scheme}. Each information exchange among followers could be established stochastically  $\forall n,m \in {\cal N}$. 
 
\subsection{Decision Making}
In this paper, the projected sub-gradient method is utilized for decision making of each follower as follows
\begin{equation}
	\begin{split}
		&x_n^{k+1}=\Pi_{{\cal X}_n}(x_n^{k}-e_n^k\alpha_n^k { g_n^k})\\
	\end{split}
	\label{follower_update}
\end{equation}
where  { $g_n^k=d_n(x_n^{k},\tilde{\sigma}_n^k,y^k)$} and  $\alpha_n^k$ is the step size of follower $n$ at iteration  $k$. { Clearly, $x_n^k$ is not updated if $e_n^k=0$.}  Also, the leader  updates its strategy at $k\in {\cal K}^L$ as follows
\begin{equation}
	\begin{split}
	y^{k+1}=\Pi_{{\cal Y}}(y^{k}-\alpha_0^k { g_0^k} );&\forall k \in {\cal K}^L\\
	y^{k+1}=y^{k};&\forall k \notin {\cal K}^L\\
	\end{split}
	\label{leader_update}
\end{equation}
where { $g_0^k=d_0(y^{k},\sigma_0^k)$} and  $\alpha_0^k$ is the step size of the leader at iteration $k$. Without loss of the generality, we set  $\alpha_0^{k+1}=\alpha_0^{k}$,  $\forall k \notin {\cal K}^L$. In this paper, the following assumptions are considered for the step sizes of the players.
\begin{assumption}
	$\alpha_n^k$, $\forall n \in {\cal N} \cup \{0\}$ are Non-increasing, $\sum_{k=0}^{\infty}\alpha_n^k=\infty $, and $\sum_{k=0}^{\infty}(\alpha_n^k)^2<\infty $.
	\label{step_size_assumption}
\end{assumption}
\begin{assumption}
	There exists $\kappa$ such that $\overline{\alpha}^k\le\kappa\underline{\alpha}^k$ where $\overline{\alpha}^k=\textbf{max}(\alpha_1^k,...,\alpha_N^k,\alpha_0^k)$ and  $\underline{\alpha}^k=\textbf{min}(\alpha_1^k,...,\alpha_N^k,\alpha_0^k)$.
	\label{different_step_size_assumption}
\end{assumption}
The optimization procedure for the leader-follower network game is presented in Algorithm \ref{optimization_communication_algorithm}.  Based on Algorithm \ref{optimization_communication_algorithm}, the leader makes decision at iterations $k^L_i \in {\cal K}^L$ and waits in other iterations, while the followers are making decision. In other words, the followers continue their interactions and decision makings based on the last informed decision of the leader for some iterations, until the next decision of the leader is announced. The initial values of the parameters are chosen from their feasible region.
\begin{algorithm}[!h]
\caption{The leader-follower network game algorithm}
\begin{algorithmic}
	\STATE \text{Initialize} $x_n$, $y$ and $\tilde{x}_{nm}$  for $\forall n,m \in  {\cal N}$ and ${k}\leftarrow 0$
	\STATE \textbf{Iteration}
	\vspace{0.05cm}
	\STATE\hspace*{0.05\linewidth}\vline
	\vspace{0.05cm} \begin{minipage}{0.95\linewidth}
		\vspace{0.05cm}
		\vspace{0.05cm}
		\STATE   \textbf{Leader}:
		\vspace{0.05cm}
		\STATE\hspace*{0.05\linewidth}
		\vspace{0.05cm}
		\begin{minipage}{0.95\linewidth}
			\STATE  $\sigma_0\leftarrow\sigma_0(x_{{\cal N}})$
			\STATE { $g_0\leftarrow d_0(y,\sigma_0)$ }
			\STATE  $y\leftarrow\Pi_{{\cal Y}}(y-\alpha_0^{k} { g_0}))$
		\end{minipage}
		\STATE \textbf{Repeat}
		\STATE\hspace*{0.05\linewidth}\vline
		\vspace{0.05cm} \begin{minipage}{0.95\linewidth}
		\STATE  \textbf{Follower} $n\in {\cal N}$:
		\vspace{0.05cm}
		\STATE\hspace*{0.05\linewidth} \begin{minipage}{0.9\linewidth}
			\vspace{0.05cm}
			\STATE \textbf{If} $e_n^k=1$:
			\STATE \hspace{0.05\linewidth}  $\tilde{\sigma}_n\leftarrow\sigma_n(\tilde{x}_{n,-n})$
			\STATE \hspace{0.05\linewidth}  { $g_n\leftarrow d_n(x_n,\tilde{\sigma}_n,y)$ }
			\STATE \hspace{0.05\linewidth} $x_n\leftarrow\Pi_{{\cal X}_n}(x_n-\alpha_n^k { g_n})$
			\vspace{0.05cm}
			\STATE  update $\tilde{x}_{nm}$  via (\ref{latest_data_follower}) based on $l_{nm}^k$   
		\end{minipage}
		\STATE ${k}\leftarrow {k}+1$
	\end{minipage}
	\STATE \textbf{Until} $k\in {\cal K}^L$
	\end{minipage}
	
\end{algorithmic}
\label{optimization_communication_algorithm}
\end{algorithm}

\section{Convergence Analysis}
\label{Convergence_Analysis_Section}
In this section, the convergence of Algorithm \ref{optimization_communication_algorithm} to the unique GNE point of ${\cal G}$ is studied. Under Assumption \ref{set_and_function_assumption}, as a result of strong convexity of the cost functions, there exists a GNE point $z^*=(x_{\cal N}^*,y^*)$ for ${\cal G}$  \cite{Rosen1965}. Before discussing the convergence, we propose the following lemmas:
{ 
	\begin{lemma}[Theorem 1 of \cite{ROBBINS1971}]	
		Let  $z_k$, $\beta_k$, $\eta_k$, and $\zeta_k$ be non-negative ${\cal F}_k$-measurable random variables. Also, assume that ${\cal F}_k$ is $\sigma$-algebra and ${\cal F}_k \subset {\cal F}_{k+1}$ holds for $\forall k \ge 0$. If  $\sum_{k=0}^{\infty}\beta_k$ and $\sum_{k=0}^{\infty}\eta_k $ almost surely converge, and the following equation
		\begin{equation}
		\textbf{E}\{z_{k+1}|{\cal F}_k\} \le (1+\beta_k)z_k + \eta_k - \zeta_k
		\label{supermartingale_eq}
		\end{equation}
		holds, then $z_k$ and $\sum_{k=0}^{\infty}\zeta_k < \infty$ almost surely converge.
		\label{supermartingale_lemma}
	\end{lemma}
}
\begin{lemma}
	$||x_n^{k+1}-x_n^k||\le A_n \alpha_n^k $   for $\forall n,m\in {\cal N}$ where $|| d_n(x_n,\sigma,y)||\le A_n, \forall x_n\in {\cal X}_n$.
	\label{iteration_increment_lemma}
\end{lemma}
\begin{IEEEproof}
	See Appendix \ref{proof_iteration_increment_lemma}.
\end{IEEEproof}
\begin{lemma}
	{ $\sum_{k=0}^{\infty}\alpha_n^k||\Delta\tilde{x}_{nm}^k|| < \infty$   for $\forall n,m\in {\cal N}$ in almost sure sense} where $\Delta\tilde{x}_{nm}^k=\tilde{x}_{nm}^k-x_m^k$.
	\label{convegence_of_the_series}
\end{lemma}
\begin{IEEEproof}
See Appendix \ref{proof_convegence_of_the_series}.
\end{IEEEproof}
\par
Using  lemmas \ref{iteration_increment_lemma} and \ref{convegence_of_the_series} , the convergence of  Algorithm \ref{optimization_communication_algorithm} is proven in Theorem \ref{convergence_theorem_different_stepsize}. 
\begin{theorem}
	Consider  Assumptions \ref{set_and_function_assumption},  \ref{probabilities_assumption}, \ref{leader_max_iteration_lenght}, \ref{step_size_assumption} and \ref{different_step_size_assumption}. If the constants $C_n$ and $C_0$ in (\ref{strong_convex_inequality}) satisfy $C_n > \frac{\kappa}{\delta}{\bar L}$ and  $C_0 > \kappa{\bar K}{\bar L}$ $\forall n\in {\cal N}$, Algorithm \ref{optimization_communication_algorithm} {almost surly} converges to the GNE point of the leader-follower game where ${\bar L}=\textbf{max}(2L,L_0)$.
	\label{convergence_theorem_different_stepsize}
\end{theorem}
\begin{IEEEproof}
	Consider the notation $\nabla x_n^k=x_n^{k} - x_n^*$. Based on Proposition 1.5.8 of \cite{Facchinei2003}, $x_n^* = \Pi_{{\cal X}_n}(x_n^* - e_n^k\alpha_n^k d_n(x_n^*,\sigma_n^*,y^*))$ where $\sigma_n^*=\sigma_n(x_{-n}^*)$. Since the projection operator $\Pi_{{\cal X}_n}(.)$ is non-expansive and $e_n^k\le 1$, we have
	\begin{equation}
	\begin{split}
	&||\nabla x_n^{k+1}||^2\le||\nabla x_n^{k}-e_n^k\alpha_n^k \big(d_n(x_n^{k},\tilde{\sigma}_n^k,y^k)\\
	&-d_n(x_n^*,\sigma_n^*,y^*)\big)||^2=||\nabla x_n^{k} ||^2 +\\
	&(e_n^k\alpha_n^k)^2 ||d_n(x_n^{k},\tilde{\sigma}_n^k,y^k)-d_n(x_n^*,\sigma_n^*,y^*)||^2\\
	&-2e_n^k\alpha_n^k \big(d_n(x_n^{k},\tilde{\sigma}_n^k,y^k)-d_n(x_n^*,\sigma_n^*,y^*)\big)^\top \nabla x_n^{k}\\
	&\le ||\nabla x_n^{k}||^2 +4A_n^2(\alpha_n^k)^2-2e_n^k\alpha_n^k\varPsi_n^k \\
	&-2e_n^k\alpha_n^k \big(d_n(x_n^{k},\tilde{\sigma}_n^k,y^k)-d_n(x_n^{k},\sigma_n^k,y^k)\big) ^\top \nabla x_n^{k}\\
	\end{split}
	\label{follower_inequality}
	\end{equation}
	where $\varPsi_n^k=\big(d_n(x_n^{k},\sigma_n^k,y^k)-d_n(x_n^*,\sigma_n^*,y^*)\big) ^\top \nabla x_n^{k}$, $\sigma_n^k=\sigma_n(x_{-n}^k)$. Let consider that $||x_1-x_2||\le B_n$ for $\forall x_1,x_2\in {\cal X}_n$. According to Assumption \ref{set_and_function_assumption}, we have
	\begin{equation}
	\begin{split}
	&-e_n^k\alpha_n^k \big(d_n(x_n^{k},\tilde{\sigma}_n^k,y^k)-d_n(x_n^{k},\sigma_n^k,y^k)\big) ^\top \nabla x_n^{k}\\
	&\le   \alpha_n^k L||\tilde{\sigma}_n^k-\sigma_n^k|| ||\nabla x_n^{k}||\le   L B_n \alpha_n^k||\tilde{\sigma}_n^k-\sigma_n^k|| \\
	& \le   L B_n \alpha_n^k \sum\limits_{m \in {\cal N}_n} w_{nm} ||\Delta\tilde{x}_{nm}^k||.\\
	\end{split} 
	\label{follower_lipschitz_inequality}
	\end{equation}
	Therefore, by putting (\ref{follower_lipschitz_inequality}) into (\ref{follower_inequality}), we have
	\begin{equation}
	\begin{split}
	&||\nabla x_n^{k+1}||^2\le ||\nabla x_n^{k}||^2 +4A_n^2(\alpha_n^k)^2 \\
	&-2e_n^k\alpha_n^k\varPsi_n^k+2 L B_n \alpha_n^k \sum\limits_{m \in {\cal N}_n} w_{nm} ||\Delta\tilde{x}_{nm}^k||\\
	&\le||\nabla x_n^{0} ||^2 +4A_n^2\sum_{k^\prime=0}^{k}(\alpha_n^{k^\prime})^2-2\sum_{k^\prime=0}^{k}e_n^{k^\prime}\alpha_n^{k^\prime}\varPsi_n^{k^\prime} \\
	&+2 L B_n  \sum\limits_{m \in {\cal N}_n} w_{nm} \sum_{k^\prime=0}^{k}\alpha_n^{k^\prime}||\Delta\tilde{x}_{nm}^{k^\prime}||\\
	\end{split} 
	\label{follower_sensivity}
	\end{equation}
	\par
	Suppose the notation $\nabla y^k=y^{k} -y^*$. Considering the leader's decision at leader's iteration $k^L_j$, and following the same operation from (\ref{follower_inequality}) to (\ref{follower_sensivity}), we have
	\begin{equation}
		\begin{split}
			&||\nabla y^{k^L_j+1}||^2\le||\nabla y^{k^L_j}||^2  +4 (\alpha_0^{k^L_j})^2 A_0^2-2\alpha_0^{k^L_j}\varPsi_0^{k^L_j}\\
			&\le||\nabla y^{0}||^2  +4\sum_{i=0}^{j} (\alpha_0^{k^L_i})^2 A_0^2-2\sum_{i=0}^{j}\alpha_0^{k^L_i}\varPsi_0^{k^L_i}.
		\end{split}
	\label{leader_inequality}
	\end{equation}
	where $\varPsi_0^{k}=\big(d_0(y^{k},\sigma_0^k)-d_0(y^{*},\sigma_0^{*})\big)^\top \nabla y^{k}$, $\sigma_0^k = \sigma_0(x_{\cal N}^k)$, $\sigma_0^* = \sigma_0(x_{\cal N}^{*})$ and $||d_0(y,{\sigma}_0)||\le A_0$. { Now, let define $\Phi^{j}=||\nabla y^{k_{j-1}^L+1} ||^2+\sum_{n\in {\cal N}}||\nabla x_n^{k_{j-1}^L+1}||^2$. Therefore, using inequalities (\ref{follower_sensivity}) and (\ref{leader_inequality}), we have:
	\begin{equation}
		\begin{split}
			&\textbf{E}\{\Phi^{j+1}\big|{\cal F}^{k^L_j}\} \le \Phi^{j}+4 (\alpha_0^{k^L_j})^2 A_0^2+4\sum\limits_{n \in {\cal N}}A_n^2\sum_{k^\prime\in {\cal K}^\prime_j}(\alpha_n^{k^\prime})^2\\
			&+ 2L \sum\limits_{n \in {\cal N}}\sum\limits_{m \in {\cal N}_n} B_n w_{nm} \sum_{k^\prime\in {\cal K}^\prime_j} \alpha_n^{k^\prime}||\Delta\tilde{x}_{nm}^{k^\prime}||\\
			&-2\alpha_0^{k^L_j}\varPsi_0^{k^L_j} -2\sum\limits_{n \in {\cal N}}\sum_{k^\prime\in {\cal K}^\prime_j} \alpha_n^{k^\prime}\textbf{E}\{e_n^{k^\prime}\big|{\cal F}^{k^\prime}\}\varPsi_n^{k^\prime}.
		\end{split}
		\label{Lyapanov_equation_raw_version}
	\end{equation}
	where ${\cal K}^\prime_j=\{k^L_{j-1}+1, \dots , k^L_j\}$.} Based on definition of $\varPsi_n^{k^\prime}$, and adding and subtracting the term $d_n(x_n^*,\sigma_n^{k^\prime},y^{k^\prime})$, we have
	\begin{equation}
	\begin{split}
	&\varPsi_n^{k^\prime}=\big(d_n(x_n^{k^\prime},\sigma_n^{k^\prime},y^{k^\prime})-d_n(x_n^*,\sigma_n^{k^\prime},y^{k^\prime})\big) ^\top \nabla x_n^{k^\prime}\\
	&+\big(d_n(x_n^*,\sigma_n^{k^\prime},y^{k^\prime})-d_n(x_n^*,\sigma_n^*,y^*)\big) ^\top \nabla x_n^{k^\prime}. \\
	\end{split}
	\label{follower_decision_with_strong_covexity_previous}
	\end{equation}
	Based on strongly convexity and Lipschitz property in Assumption \ref{strong_convex_inequality}, we have
	\begin{equation}
	\begin{split}
		&-\alpha_n^{k^\prime}e_n^{k^\prime}\varPsi_n^{k^\prime}\le -\alpha_n^{k^\prime}e_n^{k^\prime}C_n ||\nabla x_n^{k^\prime}||^2 +\alpha_n^{k^\prime} L \big( ||\nabla y^{k^L_i}||\\
		&+||\sigma_0^{k^L_i}-\sigma_0^{*}|| \big) ||\nabla x_n^{k^\prime}||. \\
	\end{split}
	\label{follower_decision_with_strong_covexity1}
	\end{equation}
	Now, using Assumptions \ref{different_step_size_assumption}, we can write \eqref{follower_decision_with_strong_covexity1} as follows
	\begin{equation}
	\begin{split}
	&-\alpha_n^{k^\prime}e_n^{k^\prime}\varPsi_n^{k^\prime}\le -\underline{\alpha}^{k^\prime} C_n e_n^{k^\prime} ||\nabla x_n^{k^\prime}||^2\\ 
	&+\overline{\alpha}^{k^\prime}\kappa  L ||\nabla x_n^{k^\prime}||(||\nabla y^{k^\prime}||+\sum_{m\in {\cal N}_n}w_{nm}||\nabla x_m^{k^\prime}||).\\
	\end{split}
	\label{follower_decision_with_strong_covexity}
	\end{equation}
	By following the same procedure for the leader, we have 
	\begin{equation}
		\begin{split}
			-\alpha_0^{k^L_j}\varPsi_0^{k^L_j}\le& -\underline{\alpha}^{k^L_j} C_0 ||\nabla y^{k^L_j}||^2\\
			& +\overline{\alpha}^{k^L_j} L_0\sum_{n\in {\cal N}}w_{0n}||\nabla x_n^{k^L_j}|| ||\nabla y^{k^L_j}||. \\
		\end{split}
	\label{leader_decision_with_strong_covexity}
	\end{equation}
	Considering Assumption \ref{probabilities_assumption}, it's clear that $\textbf{E}\{e_n^{k^\prime}\big|{\cal F}^{k^\prime}\}=q_n^k \ge \delta$. Therefore, applying inequalities (\ref{follower_decision_with_strong_covexity}) and (\ref{leader_decision_with_strong_covexity}) into the last two terms of (\ref{Lyapanov_equation_raw_version}), it can be concluded that
	\begin{equation}
		\begin{split}
			&-\underline{\alpha}^{{k^L_j}}\varPsi_0^{k^L_j}-\sum\limits_{n \in {\cal N}}\sum_{k^\prime\in {\cal K}^\prime_j} \alpha_n^{k^\prime}\textbf{E}\{e_n^{k^\prime}\big|{\cal F}^{k^\prime}\}\varPsi_n^{k^\prime}\le\\
			&-\underline{\alpha}^{{k^L_j}}C_0 ||\nabla y^{k^L_j}||^2 - \sum\limits_{n \in {\cal N}}\sum_{k^\prime\in {\cal K}^\prime_j}\delta\underline{\alpha}^{k^\prime}C_n ||\nabla x_n^{k^\prime}||^2\\
			&+\sum_{k^\prime\in {\cal K}^\prime_j}\overline{\alpha}^{k^\prime} v^{k^\prime\top}{\cal V}^{k^\prime}v^{k^\prime}.
		\end{split}
		\label{inequality_of_leader_follower_actions}
	\end{equation}
	where $v^{k^\prime}=\textbf{col}(||\nabla x_1^{k^\prime}||,...,||\nabla x_N^{k^\prime}||,||\nabla y^{k^\prime}||)$ and 
	\begin{equation*}
		{\cal V}^{k^\prime}=\left\{
		\begin{array}{lr}
		\left[\begin{array}{cc}
		L{\cal W} & L \vec{1}_N\\
		L_0\vec{w}_0^\top & 0
		\end{array} \right] &  k^{\prime}\in {\cal K}^L\\ 
		\left[\begin{array}{cc}
		L{\cal W} & L \vec{1}_N \\
		{\vec 0}_N^\top  &     0
		\end{array} \right] & k^{\prime}\notin {\cal K}^L
		\end{array}
		\right.
	\end{equation*}
	It is straightforward to see that the summation of each row of the matrices $\left[\begin{array}{cc}L{\cal W} & L \vec{1}_N\end{array} \right]$ and $\left[\begin{array}{cc}L_0\vec{w}_0^\top & 0\end{array} \right]$ are equal to $2L$ and $L_0$, respectively. Therefore, based on Perron-Frobenius Theorem \cite{Cheng_2012}, $||{\mathcal V}^{k^\prime}||\le\textbf{max}(2L,L_0)$ and $||{\mathcal V}^{k^\prime}||\le 2L$ for $k^{\prime}\in {\cal K}^L$ and  $k^{\prime}\notin {\cal K}^L$, respectively. Hence, $v^{k^\prime\top}V^{k^\prime}v^{k^\prime}\le {\bar L}||v^{k^\prime}||^2$ for $\forall k^\prime \ge 0$. Therefore, 
	\begin{equation}
		\begin{split}
			&\sum_{k^\prime\in {\cal K}^\prime_j}\underline{\alpha}^{k^\prime} v^{k^\prime\top}{\cal V}^{k^\prime}v^{k^\prime} \le \sum_{k^\prime\in {\cal K}^\prime_j} \underline{\alpha}^{k^\prime}{\bar L} ||\nabla x_n^{k^\prime}||^2\\ 
			& + \underline{\alpha}^{{k^L_j}} (k_j^L-k_{j-1}^L){\bar L} ||\nabla y^{k^L_j}||^2.
		\end{split}
		\label{inequality_of_leader_follower_actions_second}
	\end{equation}
	where the last term of \eqref{inequality_of_leader_follower_actions_second} is rearranged, since the leader only makes decision at $k^L_j \in {\cal K}^L$ and therefore, the term $||\nabla y^{k^L_j}||^2$ is the same from iteration $k_{j-1}^L$ to $k_j^L$. By putting (\ref{inequality_of_leader_follower_actions}) and  (\ref{inequality_of_leader_follower_actions_second}) into (\ref{Lyapanov_equation_raw_version}), it can be concluded that
	{\begin{equation}
		\begin{split}
			\textbf{E}\{\Phi^{j+1}\big|{\cal F}^{k^L_j}\} & \le \Phi^{j} + \underbrace{ 4 (\alpha_0^{k^L_j})^2 A_0^2+4\sum\limits_{n \in {\cal N}}A_n^2 \sum_{k^\prime\in {\cal K}^\prime_j}(\alpha_n^{k^\prime})^2}_{T_1^j}\\
			&+\underbrace{2L \sum\limits_{n \in {\cal N}}\sum\limits_{m \in {\cal N}_n} B_n w_{nm} \sum_{k^\prime\in {\cal K}^\prime_j} \alpha_n^{k^\prime}||\Delta\tilde{x}_{nm}^{k^\prime}||}_{T_2^j}\\
			&-\underbrace{2\underline{\alpha}^{{k^L_j}}(C_0-\kappa (k_j^L-k_{j-1}^L){\bar L}) ||\nabla y^{k^L_j}||^2}_{T_3^j}\\
			&-\underbrace{2\sum\limits_{n \in {\cal N}} \sum_{k^\prime\in {\cal K}^\prime_j} \underline{\alpha}^{k^\prime}(\delta C_n-\kappa {\bar L}) ||\nabla x_n^{k^\prime}||^2}_{T_4^j}.\\
		\end{split}
	\end{equation}
	\par
	Based  on Assumption \ref{step_size_assumption}, $\sum_{j=0}^{\infty}T_1^j$  is bounded. Also,  based on Lemma \ref{convegence_of_the_series} we have $\sum_{j=0}^{\infty}T_1^j < \infty$. According to Assumption \ref{leader_max_iteration_lenght}, $C_0- (k_j^L-k_{j-1}^L)\kappa{\bar L}\ge C_0- {\bar K}\kappa{\bar L}>0$ and $\delta C_n-\kappa {\bar L}>0$.   Therefore, $T_3^j$ and $T_4^j$ are positive. Now, the assumptions of Lemma \ref{supermartingale_lemma} are satisfied and as a result, and consequently, $\sum_{j=0}^{\infty} T_3^j + T_4^j < \infty$ converges almost surely. Because of positiveness of $T_3^j$ and $T_4^j$, $\sum_{j=0}^{\infty}T_3^j < \infty$ and $\sum_{j=0}^{\infty}T_4^j < \infty$ are concluded. Therefore, based on $\sum_{k=0}^{\infty}\underline{\alpha}^k\ge \frac{1}{\kappa}\sum_{k=0}^{\infty}\overline{\alpha}^k=\infty$, both $||\nabla x_n^{k^\prime}||^2$ and $||\nabla y^{k^L_i}||^2$ almost surely converge to $0$. Thus, $x_n^{k}$ and $y^{k}$ converge almost surely to $x_n^*$ and $y^*$. }
\end{IEEEproof}
{ Theorem \ref{convergence_theorem_different_stepsize} results almost sure convergence of Algorithm \ref{optimization_communication_algorithm}. Nevertheless, almost sure convergence does not generally lead to mean square convergence. Proposition \ref{mean_square_convergence} proves the convergence of Algorithm \ref{optimization_communication_algorithm} in mean square sense.
\begin{proposition}
	Under Assumptions of Theorem \ref{convergence_theorem_different_stepsize}, Algorithm \ref{optimization_communication_algorithm} converges in mean square sense.
	\label{mean_square_convergence}
\end{proposition}
\begin{IEEEproof}
	Since almost sure convergence results the convergence in distribution, the expectation of $\sum_{j=0}^{\infty}T_3^j < \infty$ and $\sum_{j=0}^{\infty}T_4^j < \infty$ converges. Therefore, $\sum_{j=0}^{\infty}\underline{\alpha}^{{k^L_j}}\textbf{E}\{||\nabla y^{k^L_j}||^2\} < \infty$ and $\sum_{k=0}^{\infty}\underline{\alpha}^{k} \textbf{E}\{||\nabla x_n^{k}||^2\} < \infty$. Hence, because of $\sum_{k=0}^{\infty}\underline{\alpha}^k > \infty$, $\textbf{E}\{||\nabla y^{k^L_j}||^2\}$ and $\textbf{E}\{||\nabla x_n^{k}||^2\}$ converges to $0$ which means that $x_n^{k}$ and $y^{k}$ converge in mean square to $x_n^*$ and $y^*$.
\end{IEEEproof}
}

{ In Theorem \ref{convergence_theorem_different_stepsize}, the convergence of the algorithm to a GNE point is studied. In the following proposition, the uniqueness of the GNE point is proven.
\begin{proposition}
	Under the assumptions of Theorem \ref{convergence_theorem_different_stepsize}, the game (\ref{Game_equation}) has a unique GNE.
	\label{uniqueness_proposition}
\end{proposition}
\begin{IEEEproof}
	Let define  $z=\textbf{col}(x_1,...,x_N, y)$ and  $z^\prime=\textbf{col}(x_1^\prime,...,x_N^\prime, y^\prime)$. Also, consider the function  
	$g(z)=\textbf{col}(d_1(x_1,\sigma_1(x_{\cal N}),y),...,d_N(x_N,\sigma_N(x_{\cal N}),y), d_0(y,\sigma_0(x_{\cal N})))$. Therefore, by following the procedure of \eqref{follower_decision_with_strong_covexity} and  \eqref{leader_decision_with_strong_covexity}, we have
	\begin{equation}
		\begin{split}
			&\Psi = (z - z^\prime)^\top (g(z) - g(z^\prime)) = \\
			& (y - y^\prime)^\top (d_0(y,\sigma_0(x_{\cal N}) - d_0(y^\prime,\sigma_0(x_{\cal N}^\prime)) \\
			& + \sum_{n\in {\cal N}} (x_n - x_n^\prime)^\top (d_n(x_n,\sigma_1(x_{\cal N}),y) - d_n(x_n^\prime,\sigma_1(x_{\cal N}^\prime),y^\prime)) \\
			&\ge C_0 ||\nabla y||^2 -  L_0\sum_{n\in {\cal N}}w_{0n}||\nabla x_n|| ||\nabla y|| \\
			& + \sum_{n\in {\cal N}} C_n ||\nabla x_n||^2 -  L \big( ||\nabla y||+\sum_{m\in {\cal N}_n}w_{nm}||\nabla x_m|| \big) ||\nabla x_n|| \\
		\end{split}
	\end{equation}
	where $\nabla x_n = x_n - x_n^\prime$ and $\nabla y = y - y^\prime$. Considering  $v=\textbf{col}(||\nabla x_1||,...,||\nabla x_N||,||\nabla y||)$, it can be concluded that
	\begin{equation*}
	\begin{split}
		&\Psi \ge C_0 ||\nabla y||^2 + \sum_{n\in {\cal N}} C_n ||\nabla x_n||^2 - v^\top {\cal R} v \\
		&  {\cal R} = {\left[ \begin{array}{cc} L {\cal W} & L \vec{1}_N\\ L_0 \vec{w}_0^\top & 0\\ \end{array} \right]}.
	\end{split}
	\end{equation*}
	Based on Perron-Frobenius Theorem, $v^\top {\cal R} v\le {\bar L} ||v||^2$. Also, it is clear that $||v||^2 = ||\nabla y||^2 + \sum_{n\in {\cal N}} ||\nabla x_n||^2$. Therefore, it can be deduced that 
	\begin{equation}
	\begin{split}
	&\Psi \ge (C_0 - {\bar L}) ||\nabla y||^2 + \sum_{n\in {\cal N}} (C_n - {\bar L}) ||\nabla x_n||^2. \\
	\end{split}
	\end{equation}
	Based on the assumptions of Theorem \ref{convergence_theorem_different_stepsize}, $C_n\ge \frac{\kappa}{\delta}{\bar L} > {\bar L}$ and  $C_0 > \kappa{\bar K}{\bar L}\ge {\bar L}$ because $\kappa, {\bar K} \ge 1$ and $\delta \le 1$. {Hence, $g(z)$ is strictly monotone  since $\Psi > 0$. Consequently, according to Theorem 2 of \cite{Rosen1965},  the GNE of the game (\ref{Game_equation}) is unique.}
\end{IEEEproof}
}
\section{Simulation Results}
\label{Simulation_Results}
As an application of leader-follower network aggregative  game, we study the power allocation of small cell networks  proposed in \cite{Semasinghe2018}. Consider a network consisting of $N$ small cells, all of which underlay a macrocell with a macrocell base station (MBS). { Small cells and macrocell provide radio coverage for cellular networks. However, small cells are low-power and have limited coverage range in comparison with macrocell.} Each small cell is considered to have a small cell base station (SBS) which can cover many users.  { Deployment of multiple SBS in a region may cause some overlapping coverage region among SBSs. In such region, transmission powers of SBSs cause the signal interference in cellular networks. As a result, based on Shanon formula utilized in (\ref{SBScost}), the data rate will be decreased. In this case, the interference appears as an aggregative term in the cost function of neighboring SBSs and therefore, the problem can be modeled as a network aggregative game. In this NAG, each SBS aims to adjust its transmission power to minimize its cost. Moreover, MBS, as a leader of small cells' network, determines the price of transmission power for SBSs as its decision variable. Consequently, there is a leader-follower NAG among the SBSs and MBS.}  Let  ${\cal N}$ denotes the set of small cells.  $x_n$ denotes the power of SBS $n\in {\cal N}$ which satisfies $0\le x_n \le {\bar P}_n$. The objective function of SBS $n\in {\cal N}$ is as follows:
\begin{equation}
\begin{split}
&J_n(x_n,x_{-n},\lambda)=R_n(S_n)-\lambda v_nx_n\\
& R_n(S_n)=A Ln(1+S_n)\\
&S_n=\frac{r_n^{-\beta}x_n}{N_0+\sum_{m\in {\cal N}_n}r_{nm}^{-\beta}x_m}, v_n=\sum_{n\in {\cal N}}r_{nm}^{-\beta}
\label{SBScost}
\end{split}
\end{equation}
where $S_n$ and $R_n(S_n)$ indicate signal to interference and noise ratio and the data rate corresponding to the SBS $n$, respectively. $A$ is the channel bandwidth and so, the first term in \eqref{SBScost} represents the transmission rate of SBS $n$. SBS $n$ has the strategy $x_n$. $r_n$ and $r_{nm}$ are the average distance of SBS $n$ to its users and the distance of SBS $n$ to SBS $m$, respectively. Based on the coverage range, SBS $n$ could be interfered from a set of other SBSs indicated by ${\cal N}_n$. $\beta$ is the path-loss exponent and $N_0$ is the white noise spectral density. The penalty term $\lambda v_nx_n$ specifies the cost for making interference to other SBSs in the network where, $\lambda$ is the penalty price. The objective function of the MBS is as follows:
\begin{equation}
\begin{split}
&J_0(\lambda,x_{\cal N})=\lambda \sum_{n\in {\cal N}} v_n x_n - B_0 \lambda ^2
\end{split}
\end{equation}
 $\lambda$ satisfies $0\le \lambda \le {\bar \lambda}$. The term $B_0 \lambda^2$ prevents MBS to increase  $\lambda$ too much. Based on the proposed framework of the game, the MBS and  SBSs can be considered as a leader and followers, respectively. 
 
 \begin{figure}[t!]
 	\centering
 	\includegraphics[width=7cm,height=5cm]{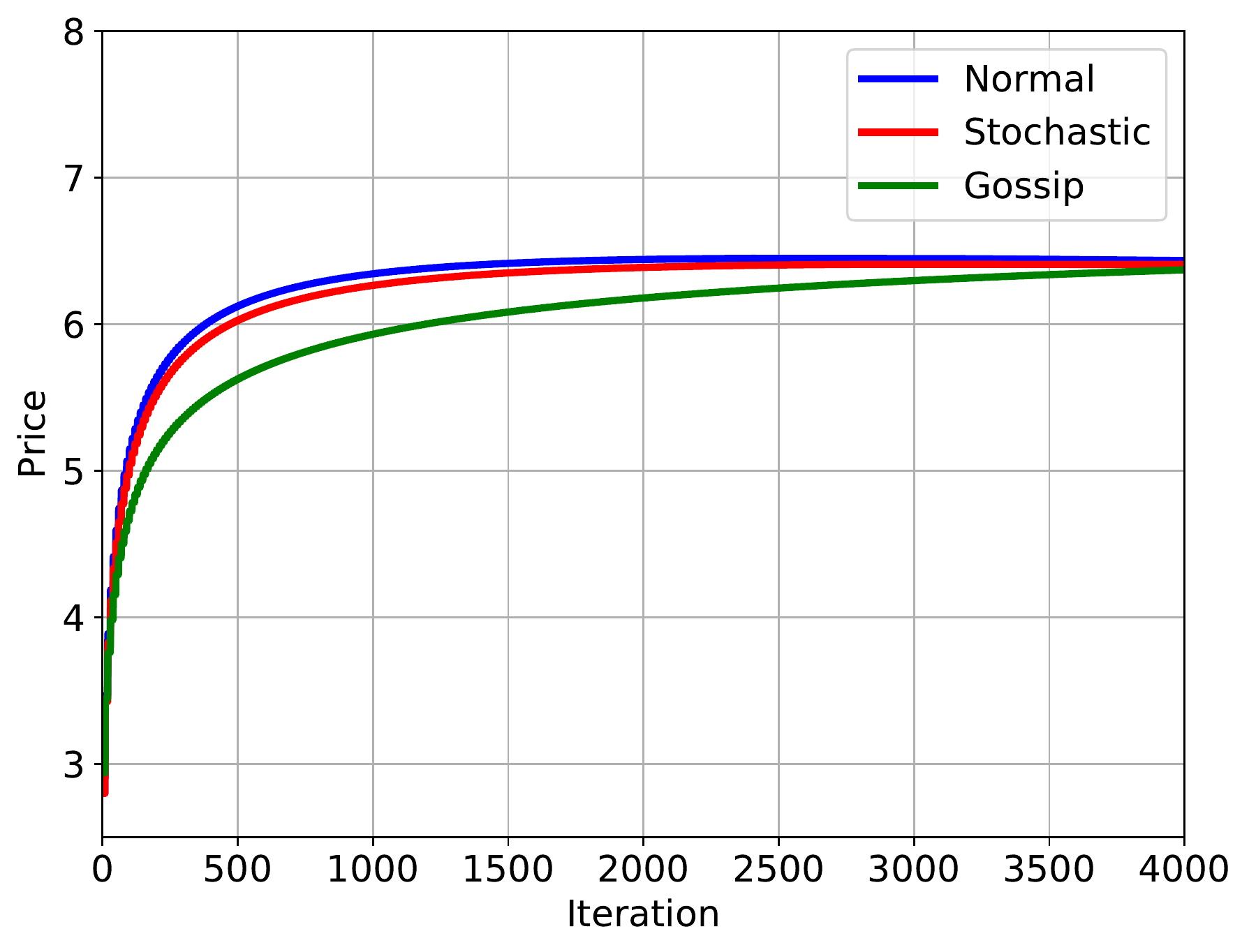}
 	\caption{The price of leader along the algorithm iterations.}
 	\label{Simulation_price}
 \end{figure}

\par
\begin{figure}[t!]
	\centering
	\includegraphics[width=7cm,height=5cm]{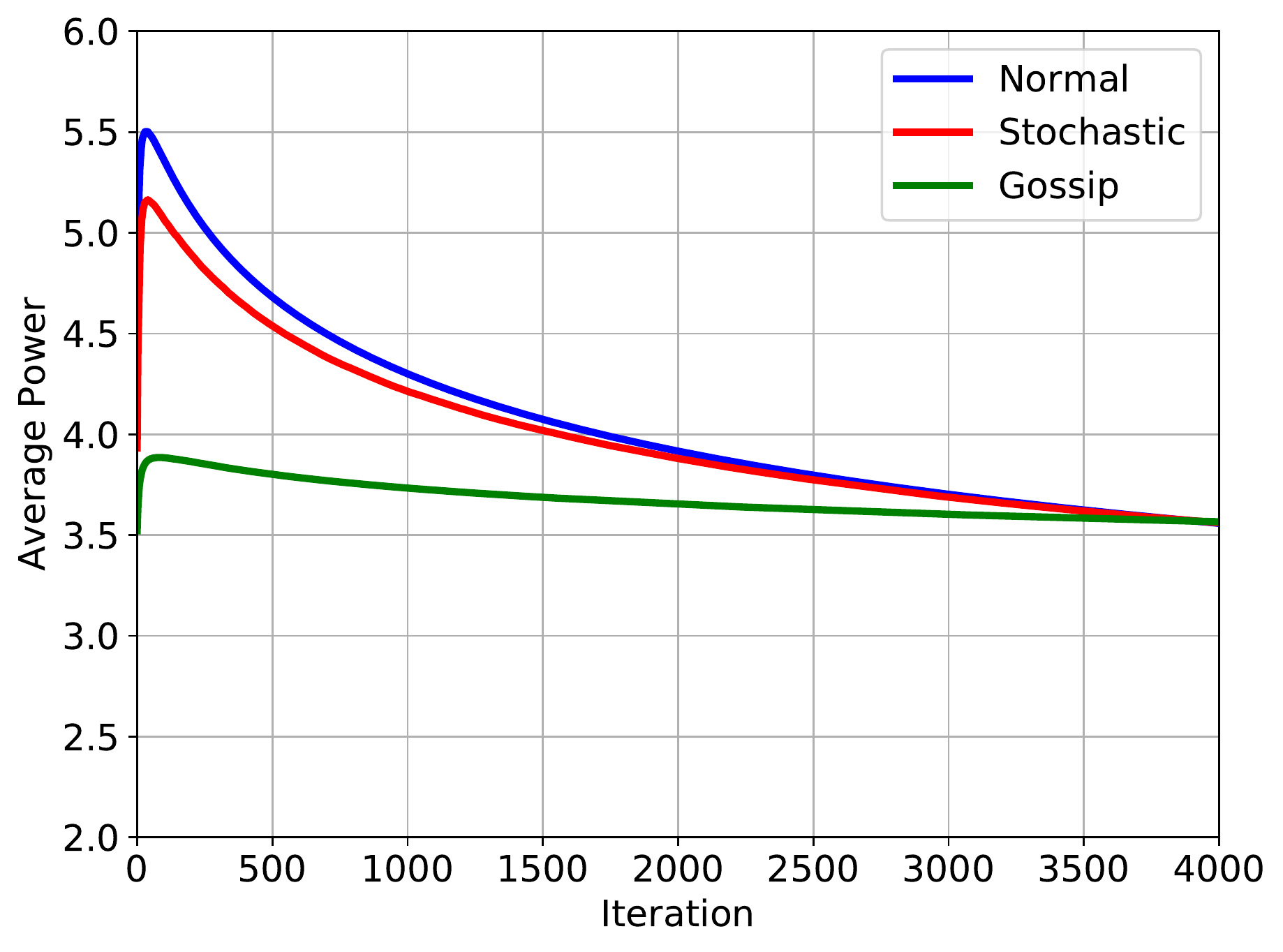}
	\caption{The average power of followers along the algorithm iterations.}
	\label{Simulation_objective}
\end{figure}

 
For simulation, 10 SBSs are considered to be stochastically located in a circular region with radius 4km. It is assumed that each of two SBSs with less than 1km distances are neighbors and have an interference effect on each other. Also, $A=2048$bps, ${\bar P}_n=6$w, $\beta=1$,  for $\forall n \in {\cal N}$. Also ${\bar \lambda}=7$ and $B_0=100$. We assume that MBS (the leader) makes decisions periodically once at every 10 iterations. The simulation is done for three communication protocols; 1) Normal ($p_{nm}^k=q_n^k=1$ for $\forall n\in {\cal N},m \in {\cal N}_n,\forall k \ge 0$) 2) Stochastic ($p_{nm}^k=q_n^k=0.7$ for $\forall n\in {\cal N},m \in {\cal N}_n,\forall k \ge 0$) 3) Gossip. The results for the leader's price and average power of the followers  are shown in Fig. \ref{Simulation_price} and Fig. \ref{Simulation_objective}, respectively. As it can be seen from Fig. \ref{Simulation_price}, the price diagram is a piecewise-constant signal based on periodic iterations of the leader. Also, as shown in Fig. \ref{Simulation_objective}, the price converges slowly in gossip-based protocol  compared to two other protocols, since just two of followers communicate and update at each iteration. Clearly, the lesser the number of updates, the slower the progress of optimization toward the equilibrium point for the followers. The stochastic scenario has an acceptable performance in comparison with Normal scenario, but with lesser active agents. Therefore, the agents can economically communicate with each other and update their decisions.

\section{Conclusion}
\label{Conclusion_FutureWork_Section}
In this paper, a leader-follower scheme was proposed for network aggregative games. Each follower was affected by both aggregated strategies of its neighbors and the leader. But, the leader was only affected by an aggregation of all followers' strategies. The leader and followers were adopted to different types of communication protocols. The leader infinitely often  became active and updated its decision and broadcasted it to the followers, nevertheless, each follower became active and communicated with its neighbors based on two different stochastic binary distributions. In particular, the aim was to find the optimal non-cooperative game solution when the agents are selfish players. A distributed optimization algorithm was proposed and it was  proven that the algorithm converges to the unique GNE point of the game in both mean square and almost sure senses. To prove the convergence of algorithm, we imposed the assumption of strong convexity to the cost functions. { There are some methods in the literature such as  Tikhonov regularization and Proximal point   \cite{Yi2019,Buong2007} which can handle the optimization problem with lower level of convexity. To this end, these methods utilize an extra quadratic term in the update rule of optimization. As a future work of this technical note, one can explore such methods to find a more relaxed condition on the cost function of the leader and followers.}

\appendix
\subsection{Proof of Lemma \ref{iteration_increment_lemma}}
\label{proof_iteration_increment_lemma}
Based on (\ref{follower_update}) and $ e_n^k\le 1$, we have 
\begin{equation*}
	\begin{split}
		&||x_n^{k+1}-x_n^{k}||=||\Pi_{{\cal X}_n}(x_n^k-e_n^k\alpha_n^k d_n(x_n^k,\tilde{\sigma}_n^k,y^k))-x_n^k ||\\
		&\le e_n^k\alpha_n^k || d_n(x_n^k,\tilde{\sigma}_n^k,y^k)||\le \alpha_n^k A_n.
		\end{split}
\end{equation*}
\subsection{Proof of Lemma \ref{convegence_of_the_series}}
\label{proof_convegence_of_the_series}
 From (\ref{latest_data_follower}) and Lemma \ref{iteration_increment_lemma}, it can be deduced that
\begin{equation}
	\begin{split}
		&\textbf{E}\{||\Delta\tilde{x}_{nm}^{k+1}||\big|{\cal F}^{k}\} = (1-\textbf{E}\{l_{nm}^k\big|{\cal F}^{k}\})  ||\tilde{x}_{nm}^{k} - x_m^{k+1}||\\
		&=(1-p_{nm}^k)  ||(\tilde{x}_{nm}^{k}-x_m^{k} )- (x_m^{k+1}-x_m^{k})||\\
		&\le(1-\gamma)  (||\Delta\tilde{x}_{nm}^{k}||+\alpha_m^{k} A_m).\\
	\end{split}
	\label{lemma_first_inequality_eq}
\end{equation}
From assumption \ref{step_size_assumption}, $\alpha_n^k$ is non-increasing, e.i. $\alpha_n^{k+1} \le \alpha_n^{k}$. Therefore, by multiplying $\alpha_n^{k+1} \le \alpha_n^{k}$ and (\ref{lemma_first_inequality_eq}), we have: 
\begin{equation}
	\begin{split}
		&\textbf{E}\{\alpha_n^{k+1}||\Delta\tilde{x}_{nm}^{k+1}||\big|{\cal F}^{k}\} \\
		&\le (1-\gamma)  (\alpha_n^{k}||\Delta\tilde{x}_{nm}^{k}||+\alpha_n^{k}\alpha_m^{k} A_m)\\
		&=\alpha_n^{k}||\Delta\tilde{x}_{nm}^{k}|| - \gamma \alpha_n^{k}||\Delta\tilde{x}_{nm}^{k}|| + (1-\gamma)\alpha_n^{k}\alpha_m^{k} A_m.
	\end{split}
	\label{lemma_communication_convergence_supermartingale_ineq}
\end{equation}
Based on Assumption \ref{step_size_assumption}, it is straightforward that  $\sum_{k=1}^{\infty}\alpha_p^{k}\alpha_m^{k}<\infty$. Consequently, the assumptions of Lemma \ref{supermartingale_lemma} are satisfied in  (\ref{lemma_communication_convergence_supermartingale_ineq}). As a result, $\sum_{k=1}^{\infty}\alpha_n^k||\Delta\tilde{x}_{nm}^k||<\infty$.

\bibliographystyle{IEEEtran}
\bibliography{MyRef}

\end{document}